\documentclass[12pt]{article}
\usepackage{times}
\usepackage{epsfig}
\usepackage{amsmath}
\usepackage{amssymb}
\usepackage{natbib}
\begin{document}

\begin{center}
{\large HCO$^+$ and radio continuum emission from the star forming region G75.78+0.34}\\
\vspace{0.5cm}
{\it Rogemar A. Riffel\footnote{e-mail address: rogemar@ufsm.br} \& Everton L\"udke\footnote{e-mail address: eludke@hotmail.com}}\\
\vspace{0.5cm}
Universidade Federal de Santa Maria, Departamento de F\'\i sica, \\ Centro de Ci\^encias Naturais e Exatas, 
97105-900, Santa Maria, RS, Brazil \\
\end{center}
\vspace{0.8cm}
\newcommand{\farcs}{\mbox{$.\!\!^{\prime\prime}$}}


\begin{abstract} 

We present 1.3 and 3.6~cm radio continuum images and a HCO$^+$ spectrum of the massive star forming region G75.78+0.34 obtained with the Very Large Array (VLA) and with the Berkley Illinois Maryland Association (BIMA) interferometer. Three structures were detected in the continuum emission: one associated to the well known cometary H~{\sc ii} region, plus two more compact structures located at 6$^{\prime\prime}$ east and at 2$^{\prime\prime}$ south of cometary H\,{\sc ii} region. Using the total flux and intensity peak we estimated an electron density of $\approx$1.5$\times$10$^{4} {\rm cm^{-3}}$, an emission measure of $\approx$6$\times$10$^{7} {\rm cm^{-6}~pc}$, a mass of ionized gas of $\approx 3\,{\rm M_\odot}$ and a diameter of 0.05\,pc for the cometary H\,{\sc ii} region, being typical values for an ultracompact H\,{\sc ii} region. The HCO$^+$ emission is probably originated from the molecular outflows previously observed in HCN and CO.
~\\
{\it Keywords}: Interstellar medium; Radio Continuum; Molecular emission; HII regions
\end{abstract}

\begin{figure}[t]
 \centering
 \includegraphics[scale=0.6,angle=-90]{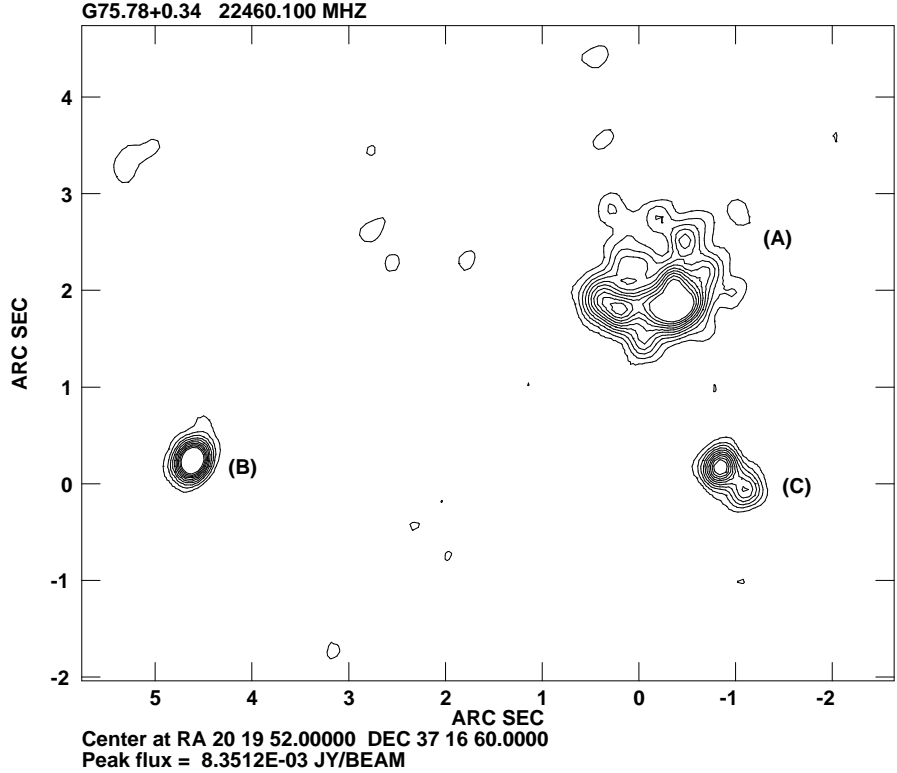}
 \caption{1.3~cm radio continuum image of G75.78+0.34 obtained with VLA at ``A" configuration. Flux levels are: 0.5, 1.0, 1.5, 2.0, 2.5, 3.0, 3.5, 4.0, 4.5 and 5.0 mJy/beam. } 
 \label{22}  
 \end{figure}

\begin{figure}[t,h]
 \centering
 \includegraphics[scale=0.85]{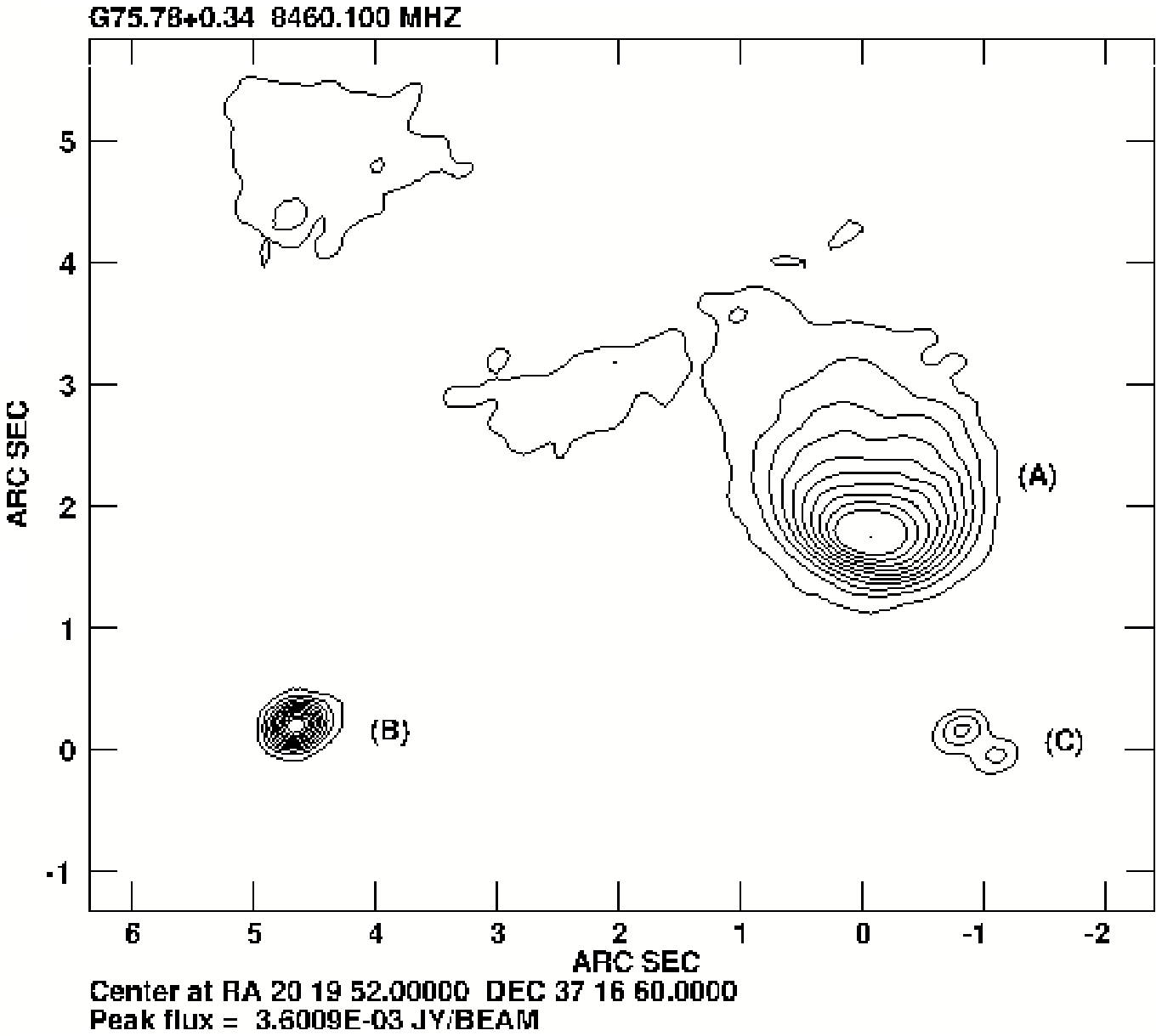}
 \caption{3.5~cm radio continuum image of G75.78+0.34 obtained with VLA at ``A" configuration. Flux levels are: 0.36, 0.72, 1.08, 1.44, 1.80, 2.16, 2.52, 2.88, 3.24 and 3.6 mJy/beam.} 
 \label{8}  
 \end{figure}

\section{Introduction} \label{intro}

H\,{\sc ii} regions are classified as ultracompact, compact 
and classical according to their sizes, ionized gas masses, densities  and
emission measures \citep[e.g.][]{habing79}. Classical H\,{\sc ii} regions such as the Orion nebulae have sizes of $\sim$10\,pc, densities of $\sim$100\,cm$^{−3}$, $\sim$10$^5$\,M$_\odot$ of ionized gas and
EM$\sim$10$^2$\,pc\,cm$^{−6}$. Compact H\,{\sc ii} regions have densities
$\gtrsim5\times$10$^3$\,cm$^{−3}$, sizes of $\lesssim$0.5\,pc, ionized gas mass of $\sim$1\,M$_\odot$ and
EM$\gtrsim$10$^7$\,pc\,cm$^{-6}$ while Ultracompact H\,{\sc ii} regions have sizes
of $\lesssim$0.1\,pc, and thermal electron densities of
$\gtrsim$10$^4$\,cm$^{−3}$, mass of ionized 
gas of $\sim$10$^{-2}$\,M$_\odot$ and EM$\gtrsim$10$^7$\,pc\,cm$^{-6}$ \citep{franco00,kurtz05}.

The study of the physical properties of H\,{\sc ii} regions and their classification is a fundamental key to understand how they evolute and how stars form.  The radio continuum emission at centimeter wavelengths traces the emission of the ionized gas. In the other hand, the study of the molecular gas next to H\,{\sc ii} regions
 provides significant information about these objects and issues of star formation theory applied to these objects. So far, most of the studies of the molecular gas are based on CO emission, which has a low dipole moment, implying that low rotational transitions do not trace dense gas, while
molecules with higher dipole moments can be used to observe high density
gas. The high density cores ($n\sim10^4 - 10^5\,{\rm~cm^{−3}}$) can be studied
from mm and submm line emission of the HCN and HCO$^+$ molecules \citep{afonso98}.

In this work, we use 1.3 and 3.6~cm Very Large Array images to study the
physical conditions of the gas in the ultracompact H\,{\sc ii} region
G75.78+0.34, as well as Berkley Illinois Maryland Association (BIMA)
interferometric data at 3 mm.  This object presents a well known molecular 
outflow observed in CO \citep{shepherd96,shepherd97} and HCN \citep{riffel}, 
is located the giant molecular cloud ON2 and was firstly identified by 
\citet{matthews73} with observations at 5 and 10.7 GHz. 
\citet{monge13} used the VLA observations to study the gas content of
G75.78+0.34. They found radio emission coming from three components: a 
cometary ultracompact  H\,{\sc ii} region, excited by a B0 type star, and 
with no associated dust emission, an almost unresolved ultracompact H\,{\sc
  ii} region, associated with a compact dust clump detected at millimetric and 
mid-infrared wavelengths and a compact source embedded in a dust
condensation. The continuum emission at 3.5~mm of G75.78+0.34 is dominated by 
free-free emission from the ionized gas surrounding the exciting star and dust 
emission may contribute only with a small fraction of its 3.5 mm 
continuum \citep{riffel}.

This paper is organized as follows. In section 2 we describe the observations, section 3 presents our results, which are discussed in section 4. The conclusions of this work are presented in section 5. 

\section{Observations}
\subsection{The BIMA data}

The observations of the emission of HCO$^+(J=1-0)$  at 89.1885 GHz and
HCN$(J=1-0)$ at 88.63\,GHz from G75.78+0.34 and G75.77+0.34 were performed
with the Berkley Illinois Maryland Association (BIMA) interferometer
\citet{welch96} in June, 1999 using its shortest baseline configuration 
(D-array).

Mars and 3C273 observations were used as primary amplitude and bandpass
calibrators and the data reduction followed standard procedures with the
software {\sc miriad} \citet{miriad95}. The  full width at half maximum
(FWHM) of the synthesized map is about 18 arcsec and the resulting velocity
sampling is $\delta V\approx 0.34\,{\rm km\,s^{-1}}$. The HCN emission has
already been presented and discussed in \citet{riffel}, where more details 
about the observations and data reduction process can be found. Here, we 
present and discuss the spectrum for the HCO$^+$.

\subsection{Radio continuum images}

We used Very Large Array (VLA) radio-continuum observations of G75.78+0.34 at
8.46~GHz (3.6 cm) and 22.46~GHz (1.3 cm) from the VLA data archive obtained
using the interferometric array at configuration ``A". These observations were
processed following the standard procedure of VLA radio continuum imaging
processing using the NRAO Astronomical Image Processing System ({\sc aips}). 
The angular resolution at these frequencies are 0\farcs24 and 0\farcs08 for 
3.6~cm and 1.3~cm observations, respectively. These resolutions corresponds 
to a spatial resolution of  $6.5\times10^{-3}$\,pc and $2.2\times10^{-3}$\,pc 
assuming a distance of 5.6\,kpc to G75.78+0.34 \citep{wood89} from which
we estimate  the sizes of the main components A, B and C are respectively 
0.065, 0.013 and 0.027 pc. 

\section{Results}

\subsection{Continuum emission}

In Figure~\ref{22} we show the 1.3 cm continuum image of G75.78+34, showing
thee components identified as A, B and C in the figure. Our image is in good
agreement with the one presented by \citet{monge13} using the same instrument
configuration. The component A show a complex structure, being more extended
to the north-south direction and presenting at least two knots of higher
emission. It is associated to the cometary H\,{\sc ii} region observed at
6\,cm by \citet{wood89} and their image appears to be smoother than
ours, probably because of its lower spatial resolution. In the other
hand, the components B and C are more compact and were not detected at
6\,cm. The component C is co-spatial with the water maser emission detected by
\citet{hofner96}. 

The 3.6~cm image of G75.78+0.34 is shown in Figure~\ref{8}, showing the same
three components observed at the 1.3~cm image. The complex structure seen 
for the component A at 1.3~cm is not seen at 3.6~cm image, because of its
lower spatial resolution.  Our 3.6~cm image is very similar to the one shown
by \citet{wood89}. The linear sizes of the components seen at 3.6~cm image are
similar to those listed above for the 1.3~cm image.  

\begin{table}
\label{fluxes}
\caption{Total flux and peak intensity for each component of G75.78+0.34 at 1.3~cm and 3.6~cm.  }
\begin{center}
\begin{tabular}{l c c c }
\hline
Component &  A  &  B&  C \\ \hline
S$_\nu$(1.3\,cm) (mJy) & 42.4 &8.6&8.2 \\ 
S$_\nu$(3.6\,cm) (mJy)& 50.4&4.7&1.6\\
 I$_\nu$(1.3\,cm) (mJy/beam)&6.7&8.4 &5.9\\
 I$_\nu$(3.6\,cm) (mJy/beam)&3.6&3.4 & 1.0 \\ \hline
\end{tabular}
\end{center}
\end{table}

Table~1 presents the measured total flux ($ S_\nu$) and the peak
intensity ($ I_\nu$) for each component identified in Figures~\ref{22} and 
\ref{8} at 1.3~cm and 3.6~cm, respectively.  
 
\subsection{HCO$^+$ line emission}

\begin{figure}[h,t]
 \centering
 \includegraphics[scale=0.65]{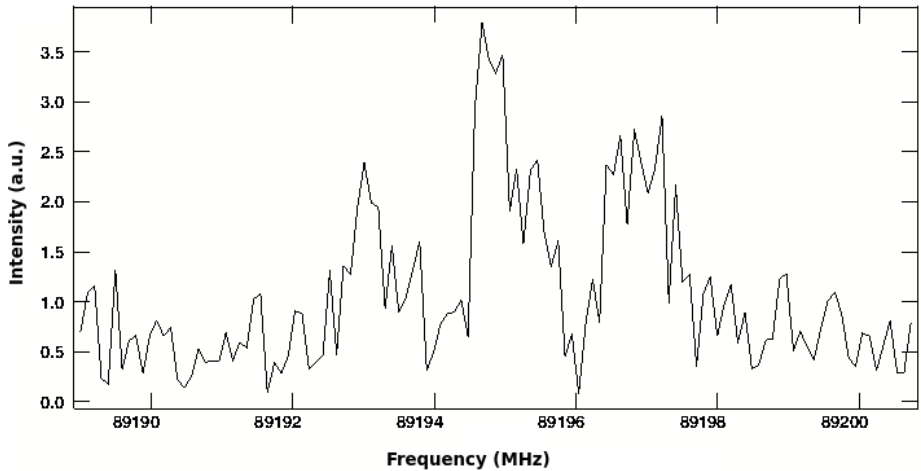}
 \caption{HCO$^+$ spectrum for G75.78+0.34 and G75.77+0.34 from BIMA
   observations. The intensities are shown in arbitrary correlator units.} 
 \label{spec}  
 \end{figure}

The signal-to-noise ratio of our BIMA observations was not high enough to construct
emission line flux maps and channel maps across the HCO$^+$ profile for BIMA
in the closest spacing configuration with hybrid self-calibration and closure
mapping, as we were able to do for the HCN in \citet{riffel} discussed as our
first results.   The HCO$^+$ spectrum is shown in figure~\ref{spec} and it was
obtained by integrating over the whole field of our BIMA observations and
include both, G75.78+0.34 and G75.77+0.34 H{\sc\,{ii}} regions. The HCO$^+$ 
emission shows three intensity peaks, at 89.193, 89.195 and 89.197 GHz, 
suggesting the presence of HCO$^+$ clouds with distinct kinematics.

\section{Discussion}

\subsection{Ionized gas}

The radio continuum images presented here can be used to investigate the
properties of the ionized gas associated to G75.78+0.34. We can use the 
total fluxes and peak intensities from Table~1 to calculate 
some physical parameters for the component A associated to the cometary H\,{\sc ii} region. 
Following \citet{panagia78} under the assumption that the H\,{\sc ii} region
has a roughly spherical geometry, we can obtain the electron density ($N_e$) by

\begin{equation}
\left(\frac{N_e}{\rm cm^{-3}}\right)=3.113\times 10^2 \left(\frac{S_\nu}{\rm{Jy}}\right)^{0.5}\left(\frac{T_e}{{\rm 10^4 K}}\right)^{0.25}
\left(\frac{D}{{\rm kpc}}\right)^{-0.5}b(\nu,T)^{-0.5}\theta_R^{-1.5},
\end{equation} 
where S$_\nu$is the total flux, $T_e$ is the electron temperature, $D$ is the distance to the object, $\theta_R$ is the angular radius in arcminutes and 
\[ b(\nu, T)=1+0.3195~{\rm log}\,\left(\frac{T_e}{\rm 10^4 K}\right)-0.2130~{\rm log}\left(\frac{\nu}{\rm 1 GHz}\right).
\]

The emission measure is given by the following equation

\begin{equation}
\left(\frac{EM}{\rm cm^{-6}~pc}\right)=5,638\times 10^4 \left(\frac{S_\nu}{{\rm
    Jy}}\right)\left(\frac{T_e}{{\rm 10^4
    K}}\right)~b(\nu,T)\theta_R^{-2}
\end{equation} 

The mass of ionized gas can be obtained by \citep{panagia78}:

\begin{equation}
\left(\frac{M}{\rm M_\odot}\right)=0.7934 \left(\frac{S_\nu}{{\rm
    Jy}}\right)^{0.5}
\left(\frac{T_e}{{\rm 10^4
    K}}\right)^{0.25}\left(\frac{D}
{{\rm kpc}}\right)^{2.5}b(\nu,T)^{-0.5}\theta_R^{1.5}(1+Y)^{-1}
\end{equation} 
where $Y$ is the abundance of He$^+$ relative to H$^+$.

The brightness temperature for a compact radio source is given by \citep[e.g.][]{wood89}
\[T_b = \frac{I_\nu 10^{-29} c^2}{2\nu^2 k \Omega_b},\]
where $\nu$ is the frequency, $I_\nu$ is the peak intensity in mJy/beam $k$ is
the Boltzmann 
constant, and $\Omega_b$ is the solid angle of the beam given by \citep[e.g.][]{rohlfs00}
\begin{equation}
\Omega_b = 1.133 ~\Theta_b^2,
\end{equation}
where $\Theta_b$  is the angular resolution of the observations in radians.

Finally, the optical depth ($\tau$) is estimated using
\begin{equation}
T_b=T_e(1-e^{-\tau}).
\end{equation}

In order to obtain the physical parameters above, we assumed typical values for the electron temperature and abundance of $T_e$=10$^4$ K and $Y$=0.07, respectively. Using the equations above and the angular resolutions of section 2, we obtain $b(\nu,T_e)=0.7121$ and $\Omega_b$ = 1.7$\times$10$^{-13}$ sr for the 1.3~cm image and $b(\nu,T_e)=0.8025$ e   $\Omega_b$ = 1.5 $\times$ 10$^{-12}$ sr for the 3.6~cm image. The radius ($\Theta_R$) of the component A can be measured directly from Figures~\ref{22} and \ref{8} and thus, we can estimate relevant physical parameters for G75.78+0.34.

\begin{table}
\caption{Physical parameters obtained for the cometary H\,{\sc ii} region G75.78+0.34 from the 1.3~cm and 3.6~cm images.}
\label{physpar}
\begin{center}
\begin{tabular}{l c c}
\hline
Parameter           & 1.3~cm  & 3.6~cm \\ \hline
$\theta_R$ (arcsec) & 0.9 & 1.2 \\
$N_e$ ${\rm (cm^{-3})}$ & $1.7 \times 10^4$& $1.2 \times 10^4$\\
$EM$  (${\rm cm^{-6}~pc}$) & $7.6 \times 10^6$ &$5.1 \times 10^6$\\
$M$ (${\rm M_\odot}$)& $2.5 \times 10^{-2}$&$3.9 \times 10^{-2}$\\
$T_b$ (K)& 2530&1100\\
$\tau$ & 0.29&0.12\\
Diameter (pc)& 0.05&0.06\\ \hline
\end{tabular}
\end{center}
\end{table}

In Table~2 we present the physical parameters estimated by the equations above, which can be compared with previous estimates from the literature. The values obtained for the electron density are in reasonable agreement with those obtained by \citet{wood89} from the 6~cm radio emission ($N_e=2.7\times10^4 {\rm cm^{-3}}$) and with the value obtained from 7\,mm observations -- $N_e\approx5\times10^4 {\rm cm^{-3}}$ \citep{carral97}, as well as with the one found by \citet{monge13} from a multi-frequency study ($N_e=3.7\times10^4 {\rm cm^{-3}}$). The emission measure found here is between the values found by \citet{matthews86} of $EM=1.1\times10^6 {\rm cm^{-6}~pc}$, by \citet{wood89} of $EM=2.3\times10^6 {\rm cm^{-6}~pc}$ and the one from \citet{monge13} of $EM=2.5\times10^7 {\rm cm^{-6}~pc}$. \citet{wood89} found $T_b=2300\,{\rm K}$ and $\tau=0.27$ and thus, our values obtained from the 1.3~cm emission are in good agreement with theirs, while those estimated from the 3.6~cm are a bit smaller. 
The mass of G75.78+0.34 obtained here is about one order of magnitude smaller than the value found by \citet{matthews86} and about one order of magnitude larger than the value obtained by \citet{monge13}, probably due to differences in the size of the region used to integrate the flux by these authors and by us, as well as due to the assumptions used in the calculations. The parameters presented in table~\ref{physpar} confirm that G75.78+0.34 is an ultracompact H\,{\sc ii} region \citep{kurtz05}, confirming previous results of \citet{monge13}.

For the components B and C, we estimate the physical parameters using only the
1.3~cm image, since it has the highest spatial resolution. For the component B, we obtain $N_e\approx3.3\times10^5$\,cm$^{-3}$, $EM\approx1.9\times10^7 {\rm cm^{-6}~pc}$ and a mass of ionized gas of $M\approx1.6\times10^{-3}$~M$_\odot$, assuming a radius of 0.25 arcsec for the region. Assuming a radius of 0.4 arcsec for the component C, we obtain $N_e\approx1.6\times10^5$\,cm$^{-3}$, $EM\approx7.4\times10^6 {\rm cm^{-6}~pc}$ and $M\approx3\times10^{-3}$~M$_\odot$. These values are similar to those espected for ultracompact H\,{\sc ii} regions \citep{franco00,kurtz05}. In fact, \citet{monge13} identified the component C as being an ultracompact H\,{\sc ii} region associated to dust emission in the millimetric and mid-infrared wavelengths, while the component B seems to be embedded in dust. 

\subsection{Molecular gas}

Although, the signal-to-noise ratio of the HCO$^+$ data was not high
enough to construct flux and channel maps, as it has been done for the HCN in
\citet{riffel} using the same data, its detection can tell us some information
about the star-forming region G75.78+0.34. The complex line profile shown in
Fig.~\ref{spec} suggests that the HCO$^+$ emission is originated from gas with
a disturbed kinematics and since it is usually originated from high density
gas, the HCO$^+$ emission may be tracing also the bipolar molecular outflows
observed in HCN for G75.78+0.34 in our previous paper. A similar conclusion 
has been presented for the Serpens star-forming region using BIMA observations 
by \citet{serpens}, for which it is found that the HCN and HCO$^+$ present 
enhanced emission near outflows, while the N$_2$H$^+$ reflects the
distribution of the cloud. 

\subsubsection{SEST observations and the non detection of SiO emission}

Besides the BIMA data, we have also employed the Swedish-ESO Submillimeter
Telescope (SEST)  in 20th April 2003 in order to detect higher excitation
lines like SiO (2-1 at 86.646 GHz), SiO (3-2 at 130.268 GHz) and SiO (5-4 
at 217.104 GHz) and SiO (8-7, 347.330 GHz) to make a spectral line survey 
of maser emission from G75.78+0.34 and G75.77+0.34 at higher excitations. 
Even with longer integration times of about one hour per transition, we were 
unable to detect these lines at a confidence better than $5\sigma$ above the 
instrumental noise level.  

On the other hand, the H$^{13}$CN at 86.338 GHz was present in our
SEST observations, presenting a similar profile than the one of the HCN (1-0) 
suggesting that the emission may occur within zones of similar physical and
kinematical conditions. Our observations therefore suggest that SiO lines 
may not be good observational tracers of shock-driven outflows 
in G75.78+0.34. 

This conclusion is supported in a similar result obtained for the
star-forming core W3-SE using the Combined Array for Research in Millimeter-wave Astronomy (CARMA)  
by \citet{W3}, in which the authors report the detection of the HCO$^+$ associated to outflows,
while the SiO (2-1) line at 86.243 GHz was also not detected for W3-SE in 
independent observations. This certainly brings the fact that there are 
difficulties in using SiO to obtain information on the astrophysical 
conditions of this compact object within the production of SiO by destruction 
of dust grains by radiatively-driven MHD shock front from young and massive
stars.   

On the other hand, we must emphasize that there are several works for 
UCH{\,\sc ii} regions and more evolved objects showing that the SiO is a 
good tracer of outflows \citep[e.g.][]{kumar04,minh10,zapata08} and a possible explanation for the absence of 
excited SiO lines in G75.78+0.34 could be due to low sensitivity of the 
SEST observations. New and more sensitivity observations are henceforth 
needed to better constrain the SiO emission or its absence in G75.78+0.34 
and to unravel such apparently contradictory statements.

\section{Conclusions}

We studied the radio continuum and molecular emission from the star forming
region G75.78+0.34 using VLA and BIMA observations. Our main conclusions are: 

\begin{itemize}

\item The 3.6 and 1.3~cm radio continuum images for the H\,{\sc ii} region G75.78+0.34 present three components, one associated to the cometary H\,{\sc ii} region, other located at 6$^{\prime\prime}$ east of the cometary H\,{\sc ii} region and another structure located at 2$^{\prime\prime}$ south of it.

\item The cometary H\,{\sc ii} region present an electron density of $\approx$1.5$\times$10$^{4} {\rm cm^{-3}}$, an emission measure of $\approx$6$\times$10$^{7} {\rm cm^{-6}~pc}$, ionized gas of $\approx 3\,{\rm M_\odot}$ and a diameter of 0.05\,pc, consistent with the values expected for an Ultracompact H\,{\sc ii} region.

\item The HCO$^+$ (J=1-0) seems to be originated from the gas which 
  follows the same astrophysical conditions of outflows previously observed
  in main HCN transitions. Nevertheless, new
  interferometric observations with enhanced sensitivity and resolution are
  needed to better constraint the HCO$^+$ origin in G75.78+0.34.

\item The SiO higher transitions up to 200 GHz were not detected using 
single dish observations with SEST of G75.78+0.34, suggesting that either 
it is not a good tracer of outflows in dense cores or that more sensitivity is
needed at those relevant frequencies. 
\end{itemize}

\paragraph {Acknowledgments:}
We thank the referee for valuable suggestions that helped to improve the present paper.
E. L\"udke particularly would like to thank the BIMA Staff 
and the SEST Observing Team at ESO La Silla for comments which improved the 
observations. The SEST Radiotelescope was operated by Onsala Radio 
Observatory/Sweden. The Very Large Array is a facility operated by the 
National Radio Astronomy Observatory under cooperative agreement by 
Associated Universities Inc. and the National Science Foundation. 

The authors have been partially supported by the Brazilian institutions CNPq,
CAPES and FAPERGS. 

\paragraph {Conflict of Interests:}

The authors declare that there is no conflict of interests
regarding the publication of this article.


\begin{thebibliography}{10}



\bibitem[Afonso, Yun \& Clemens(1998)]{afonso98} Afonso, J. M., Yun, J. L., \& Clemens, D. P., 1998, AJ, 115, 1111

\bibitem[Carral et al.(1997)]{carral97} Carral, P., Kurtz, S. E., Rodr\'iguez, L. F., De Pree, C., \& Hofner, P., 1997, ApJ, 486, L103.

\bibitem[Franco et al.(2000)]{franco00} Franco, J., Kurtz, S.,  Garc\'\i a-Segura, G., \& Hofner, P., 2000, ApSS, 272, 179.

\bibitem[Habing \& Israel(1979)]{habing79} Habing, H. J., Israel, F. P., 1979, ARAA, 17, 345.

		
\bibitem[Hogerheijde(2002)]{serpens} Hogerheijde, M. R. 2003, {\it SFChem 2002: Chemistry as a Diagnostic of Star Formation, proceedings of a conference held August 21-23, 2002 at University of Waterloo, Waterloo, Ontario, Canada N2L 3G1. Edited by Charles L. Curry and Michel Fich. NRC Press, Ottawa, Canada,} 305.


\bibitem[Hofner \& Churchwell(1996)]{hofner96} Hofner, P., \& Churchwell, E., 1996, A\&AS, 120, 286.

\bibitem[Kumar, Tafalla \& Bachller(2004)]{kumar04}  Kumar, M. S. N., Tafalla, M. \& Bachller, R., 2004, A\&A, 426, 195.

\bibitem[Kurtz(2005)]{kurtz05} Kurtz, S., 2005, Proceedings IAU Symposium 227, Massive Star Birth: A Crossroads of Astrophysics, R. Cesaroni, M. Felli, E., Churchwell \& C. M. Walmsley, eds. pg. 111.

\bibitem[Matthews et al.(1973)]{matthews73} Matthews, H. E., Goss, W. M., Winnberg, A., Habing, H. J., 1973, A\&A, 29, 309.

\bibitem[Matthews, Andersson \& Macdonald(1986)]{matthews86} Matthews, N., Andersson, M., Macdonald, G. H., 1986, A\&A, 155, 99.

\bibitem[Minh et al.(2010)]{minh10} Minh, Y. C., Su, Y.-N., Chen, H.-R., Liu, S.-Y., Yan, C.-H., Kim, S.-J., 2010, ApJ,  723, 1231.

\bibitem[Panagia \& Walmsley(1978)]{panagia78} Panagia, N. \& Walmsey, C. M.,1978, A\&A, 70, 711.

\bibitem[Riffel \& L\"udke(2010)]{riffel} Riffel, R. A., \& L\"udke, E., 2010, MNRAS, 404, 1449.

\bibitem[Rohlfs \& Wilson(2000)]{rohlfs00} Rohlfs, K., \& Wilson, T. L., 2000, {\it Tools of Radio Astronomy} Springer-Verlag.

\bibitem[Sault, Teuben \& Wright (1995)]{miriad95} Sault, R.J., 
Teuben, P.J. \& Wright, M.C.H., 1995, In Astronomical Data Analysis Software and Systems IV, ed. R. Shaw, H.E. Payne, J.J.E. Hayes, ASP Conference Series, 77, 433-436. 

\bibitem[S\'anchez-Monge et al.(2013)]{monge13} S\'anchez-Monge, A., Kurtz, S., Palau, A., Estalella, R., Shepherd, D., Lizano, S., Franco, J., Garay, G., 2013, ApJ, 766, 114.

\bibitem[Shepherd \& Churchwell(1996)]{shepherd96} Shepherd, D. S., \& Churchwell, E., 1996, ApJ, 472, 225.

\bibitem[Shepherd, Churchwell \& Wilner(1997)]{shepherd97} Shepherd, D. S., Churchwell, E., \& Wilner, D. J., 1997, ApJ, 482, 355.

\bibitem[Zapata et al.(2008)]{zapata08} Zapata, L. A., Leurini, S., Menten, K. M., Schilke, P., Rolffs, R., Hieret, C.,  2008, A\&A, 479, 25.

\bibitem[Zhu et al.(2010)]{W3} Zhu, L., Wright, M.C.H., Zhao, J., Wu, Y, 2010, ApJ, 712, 674.

\bibitem[Welch et al.(1996]{welch96} Welch W. J. et al., 1996, PASP, 108, 93.

\bibitem[Wood \& Churchwell(1989)]{wood89} Wood, D. O. S., \& Churchwell, E., 1989, ApJs, 69, 831.


\end{thebibliography}
\end{document}